\documentclass{rmaa}

\usepackage{paralist}
\usepackage{psfrag,color}
\usepackage[latin1]{inputenc}

\title{Evidence of Supermassive Black Holes in Narrow Emission Line Galaxies}

\author{
   J.~P. Torres-Papaqui,\altaffilmark{1}
   R. Coziol,\altaffilmark{1}
   H. Andernach\altaffilmark{1}
   R.~A. Ortega-Minakata,\altaffilmark{1}
   D.~M. Neri-Larios,\altaffilmark{1}
  and I. Plauchu-Frayn,\altaffilmark{2}}

\altaffiltext{1}{Departamento de Astronom\'{\i}a, Universidad de
Guanajuato, Guanajuato, M\'exico}

\altaffiltext{2}{Instituto de Astrof\'{\i}sica de Andaluc\'{\i}a
(CSIC), Granada, Spain}

\shortauthor{Torres-Papaqui et al.}

\shorttitle{SMBH in NELGs}

\fulladdresses{
\item H. Andernach, R. Coziol, D.~M. Neri-Larios, R.~A.
Ortega-Minakata, J.~P. Torres-Papaqui: Departamento de
Astronom\'{\i}a, Universidad de Guanajuato, Apartado Postal 144,
36000 Guanajuato, Gto, Mexico
  (heinz, rcoziol, daniel, rene, papaqui@astro.ugto.mx).
\item
I. Plauchu-Frayn: Instituto de Astrof\'{\i}sica de Andaluc\'{\i}a
(CSIC), E-18008, Granada, Spain
  (ilse@iaa.ee).}

\listofauthors{J.~P. Torres-Papaqui, R. Coziol,  H. Andernach, 
R.~A. Ortega-Minakata,\altaffilmark, D.~M.
Neri-Larios \& I. Plauchu-Frayn}

  \indexauthor{Torres-Papaqui, J.~P.}
  \indexauthor{Coziol, R.}
  \indexauthor{Andernach, H.}
  \indexauthor{Ortega-Minakata, R.~A.}
  \indexauthor{Neri-Larios, D.~M.}
  \indexauthor{Plauchu-Frayn, I.}

\abstract{A sample of 229618 narrow emission-line galaxies is used to establish two new unambiguous type of evidence for supermassive black holes at the center of their nuclei: 1) the Seyfert 2 galaxies and LINERs follow the same characteristic power law relating the luminosity of ionized flux with that of the continuum; 2)  both show the highest concentration of mass at their center, independent of the morphology of the galaxy, consistent with  higher binding energies. The Full Width at Half Maximum is shown to be related with the mass concentration, suggesting that the kinetic energy of the gas in AGNs has a gravitational origin.  Within the standard accretion model, the Transition-type Objects, Seyfert 2 galaxies and LINERs represent AGNs forming supermassive black holes on different mass-scales, or they could be related through an evolutionary process, the LINERs representing the end product of this evolution.  
}

\resumen{Una muestra de 229618 galaxias con l\'ineas angostas de emisi\'on es utilizada para establecer dos nuevos tipos de evidencias inambiguas de agujeros negros supermasivos en el centro de sus n\'ucleos: 1) las galaxias Seyfert 2 y los LINERs siguen la misma ley de potencia caracter\'istica relacionando la luminosidad del flujo ionizante con la del continuo; 2) ambas muestran la más alta concentraci\'on de masa en sus centro, independientemente de la morfolog\'ia de la galaxia, consistente con altas energ\'ias de ligaci\'on. El ancho a potencia media est\'a relacionado con la concentraci\'on de masa, sugiriendo que la energ\'ia cin\'etica del gas en los AGNs tiene un origen gravitacional. Dentro del modelo est\'andar de acreci\'on, los Objetos de Transici\'on, las galaxias Seyfert 2 y los LINERs representan AGNs formando agujeros negros supermasivos a diferentes escalas en masa, o prodr\'ian estar relacionados dentro de un proceso de evoluci\'on, donde los LINERs ser\'{\i}an el producto final.
}

\addkeyword{galaxies: activity}
\addkeyword{(galaxies:) quasars: general}
\addkeyword{galaxies: morphology}
\addkeyword{galaxies: accretion rate}

\begin{document}
% Typeset article header
\maketitle

\section{Introduction}
\label{sec:intro}

In the early 1960's, numerous unresolved radio sources were found that are now known as quasars \citep{Weedman86}. The optical spectra of quasars show they are extremely rich in hot gas, ionized by copious amounts of high-energy photons \citep{Osterbrock89,BNW90,Krolik99}. Very rapidly it was discovered that there are much more radio-quiet quasars than radio-loud ones \citep{Sandage65,Braccesi80,Schmidt83,Padovani93,deVries06}. Understanding and explaining the optical spectra of quasars was also instrumental in confirming the interpretation of the redshift in terms of the expansion of the universe \citep{Schmidt63,Heckman84}. Adopting this cosmology and putting all the observations together, with time it became clear that the energy produced by a quasar is so phenomenal that the only physical mechanism known that could explain it is the transformation of gravitational energy into radiation through the accretion of matter on a supermassive black hole (SMBH) forming at the center of galaxies \citep{Rees78,Soltan82,Netzer85}. Today, this interpretation constitutes the basis of the standard model of the Active Galactic Nuclei (AGN) phenomenon.

Subsequent observations in the 1970's and 1980's rapidly showed that quasars were much more common at high redshift, that is, in the past, which is now largely accepted as evidence of luminosity or/and density evolution with time \citep{Schmidt72,Braccesi80,Veron86,Yee87,Boyle00,Richards06,Croom09}. The common interpretation for this evolution is that luminous AGNs mark the first phase in the formation of galaxies by mergers, possibly following, very tightly and in a complex intricate way, the formation of their bulges \citep{Cavaliere86,Silk98,Miller03,DiMatteo05,Lapi06,Hopkins07a,Horst07,Elbaz09,Letawe10,Treister10}. Consistent with this interpretation it was found that the mass of the SMBH is well correlated with the bulge mass or the stellar velocity dispersion of the galaxies \citep{Magorrian98,Ferrarese00,Gebhardt00,Haring04,Peterson05,Gultekin09,Graham11}.

So far, the standard model for AGN seems to offer an elegant and credible explanation for quasars, in good agreement with cosmology. However, here lies one mystery: if one counts the number of luminous AGNs at high redshift, then almost all massive galaxies in the nearby universe must have had a comparable high-activity phase during their formation \citep{Kormendy95,Richstone98}. But what is the evidence at low redshift in favor of such a conclusion? In other words, where are the luminous quasars, or rather their remnants, at low redshift \citep{Lopez-Corredoira11}? Within this interpretation, the luminous AGN phase is assumed to be very short in duration ($10^6$ to $10^8$ yrs) compared to the formation and evolution of galaxies \citep{Martini04,Kelly10}. This suggests that most quasar remnants today should not show any trace of non-stellar activity in their nucleus. But how short really is this phase \citep{Cattaneo01}, and how consistent is this scenario with the low level of nuclear activity observed in galaxy surveys at low redshift \citep{Ueda03,Merloni04,Heckman04,Barger05}?

In the present study we use one of the largest spectroscopic survey to date, the Sloan Digital Sky Survey \citep[SDSS;][]{York00,Stoughton02}, in order to shed some light on the evolution of AGNs at low luminosity. The SDSS spectral survey has revealed that a high fraction of nearby galaxies possess ionized gas in their nucleus. A small fraction of these galaxies, the Seyfert~1 (Sy1), show broad emission-line components akin to what is observed in quasars \citep{Weedman86,Dibai87,Osterbrock89,Krolik99}. The standard model of AGN suggests that the Sy1s are similar to quasars, forming the low-luminosity tail of their distribution \citep{Smith86,Marshall87}. In terms of the SMBH theory it is expected that these AGNs have lower-mass black holes for the same accretion luminosity, $L_{acc}$, and efficiency, $\eta$, than the more luminous AGNs \citep{Hopkins06a}. 

However,  much more numerous in the SDSS survey are narrow emis\-sion-line galaxies (NELGs). Using diagnostic diagrams that compare various emis\-sion-line ratios \citep{Baldwin81,Veilleux87}, and different classification criteria \citep{Kewley01,Kauffmann03}, a good fraction of these NELGs were found to have spectra consistent with that of an AGN, although they do not show as broad lines as the Sy1 and are less luminous. One of these AGNs, the Seyfert~2 (Sy2), is thought to differ from the Sy1s only because the broad-line regions in these galaxies are obscured by a circumnuclear torus of gas and dust \citep[this is the so-called unified AGN model;][]{Antonucci85}. On the other hand, the Sy2s may also be intrinsically different from the Sy1s, lacking, for some unknown reason, the broad-line regions.

Another type of AGN which is even more problematic than Sy2 is the Low Ionization Nuclear Emission-line Regions \citep[LINER;][]{Heckman80,Coziol96,Kewley06}. Despite the classification based on standard spectral diagnostic diagrams, which identify these galaxies as AGNs similar to Sy2s, many researchers today claim that LINERs are ``false AGNs'', mostly because many show active and intense star-formation activity in their circumnuclear regions, and do not present clear evidence for a compact source in their nucleus. The presence of star formation in narrow-line AGNs seems now indubitable \citep{Kauffmann03}. This is why the Transition-type Object (TO) class was systematically included, defining in diganostic diagrams a sort of ``buffer'' zone between pure Star Forming Galaxies (SFGs) and pure (or rather dominant) AGNs. The presumed absence of a compact source in LINERs, on the other hand, was never proven.  Instead, evidence for an AGN contribution is usually downgraded in favor of star formation, but never eliminated completely \citep[e.g.,][]{Annibali10}. 

Finding evidence of SMBHs genuine AGNs should not pose such a problem. For example, in one of his books, \citet{Osterbrock89} proposed a simple test, which consists in comparing the luminosity of the emission line, say H$\alpha$, with the continuum luminosity in the blue, say at 4800\AA\ (sufficiently far from H$\beta$ at 4861\AA). According to the standard model of AGN, the source of the ionizing flux is the UV continuum produced by the accretion of matter onto a SMBH. Consequently, a typical AGN must show a characteristic power law relation, $L_{em}\propto L^{\alpha}_{cont}$, significantly different from any relation followed by SFGs (the same principle, for the same reasons, is at the basis of the spectral diagnostic diagrams). Up to now, however, this test was never applied to a large and statistically significant sample of NELGs \citep[e.g.,][]{Buson93,Macchetto96}. This is one of the major goals of our study.

The plan of this article is the following. In Section~2, we describe the selection of our sample of NELGs and explain how the data for our analysis were obtained. In Section~3 we present the results of our spectral classification analysis. In Section~4 we study and discuss the results of the Osterbrock power-law test and in Section~5 we present other evidence consistent with SMBHs at the center of NELGs. Finally, in Section~6 we discuss how the standard accretion model can explain the different types of AGNs. Our conclusions can be found in Section~7.

\section{Description of the samples and data for analysis}
\label{sec:sample}

The data for our study come from the main catalogue of the Sloan Digital Sky Survey, Data Release~7 \citep[SDSS, DR~7;][]{AM07}. Using the Virtual Observatory (VO) service\footnote{http://www.starlight.ufsc.br}, we downloaded the spectra of 229618 galaxies, all classified as NELGs, and with the highest spectral quality, namely emission lines with signal-to-noise ratio $S/N \geq$ 3 on H$\beta$, [OIII]$\lambda$5007\AA, [NII]$\lambda$6584\AA\ and H$\alpha$ and $S/N \geq$ 10 for the adjacent continuum. To minimize the Malmquist bias effect (and avoid problems with cosmology), we have limited our sample to a redshift z $\leq$ 0.25.

In the VO, the spectra are corrected for Galactic extinction, shifted to their respective rest frame, resampled to $\Delta \lambda$ = 1\AA\ between 3400 and 8900\AA,  and processed by the spectral synthesis code STARLIGHT \citep{CidFernandes05}. The STARLIGHT code produces automatically a stellar template from which important characteristics of the NELG hosts can be deduced.  In particular for our study, we retrieved from the templates the velocity dispersion of the stars, $\sigma_{ap}$, in the bulge of the galaxies. It is important to mention that STARLIGHT does not allow to correct for rotation. However, this problem is reduced in our study because the small aperture of the SDSS fiber (3 arcseconds) projected at the distance of the galaxies always covers a region well within the bulge, which is pressure supported \citep{BBF92}. We have also verified that within the resolution of the stellar libraries used in our study $\sigma_{ap}$ does not change significantly as the aperture increases. The velocity dispersion estimated by STARLIGHT converges rapidly to the real stellar velocity dispersion of the bulge, $\sigma_{\star}$. This is consistent with the study of \citet{Bernardi03a}, who have shown that the $\sigma_{\star}$ measured using SDSS spectra have an uncertainty smaller than 14\%, which is equal to the intrinsic variance of the velocity dispersion as measured in galaxies having the same morphology and redshift.  Therefore, using the effective radius, $R_e$, and $\sigma_{ap}$ the mass of the bulge of the galaxy in our sample can be reasonably estimated by applying the virial theorem. Once the bulge mass is known, we can infer the mass of the SMBH at the center of the AGNs in our sample \citep[e.g.,][]{Haring04}. However, for our study we decided to use instead the relation between M$_{BH}$ and $\sigma_{\star}$  \citep{Gultekin09,Graham11}, which is more direct because it does not require determining the effective radii of the galaxies. 

To separate the NELGs in our sample according to their different nuclear activity, we used three different spectral-line diagnostic diagrams. The line ratios necessary for the classification were measured in the corrected spectra, obtained by subtracting the stellar templates produced by STARLIGHT from the raw spectra. This method improves greatly the classification process by reducing the diluting effect on emission-lines of the underlying older stellar populations. On the other hand, some uncertainties are introduced by this method because STARLIGHT assumes the continuum is purely stellar.  However, this affects more the characteristics of the stellar templates than the emission line ratios.  That is, in AGNs the velocity dispersion and the ages of the stellar population may be slightly underestimated. Note however that to apply the Osterbrock test, the luminosity of H$\alpha$ is determined using the corrected spectra, which is free of stellar contribution, while the luminosity of the continuum, at any wavelength (we used two in our study), is measured in the raw spectra, which include the contribution of stars and the AGN if present. 

To verify if the power laws in our test depend on the morphologies, we have also determined this parameter for all the galaxies in our sample. To do so, we have followed the method developed by \citet{Shimasaku01} and \citet{Fukugita07}. Using the photometric colors, $u-g$, $g-r$, $r-i$, and $i-z$, as defined in the $ugriz$ photometric system of SDSS\footnote{http://casjobs.sdss.org}, and the inverse concentration indices, $R_{50}(r)/R_{90}(r)$, which is the ratio of the Petrosian radii \citep{Petrosian76} containing 50\% and 90\% of the total flux in the $r$ band, correlations were established with the standard Hubble morphological types \citep{deVauculeurs91}. Following \citet{Blanton07}, a K-correction was applied on the magnitudes.

\section{Nuclear activity classification}
\label{sec:classification}

\begin{figure}[!h]
  \includegraphics[width=\columnwidth]{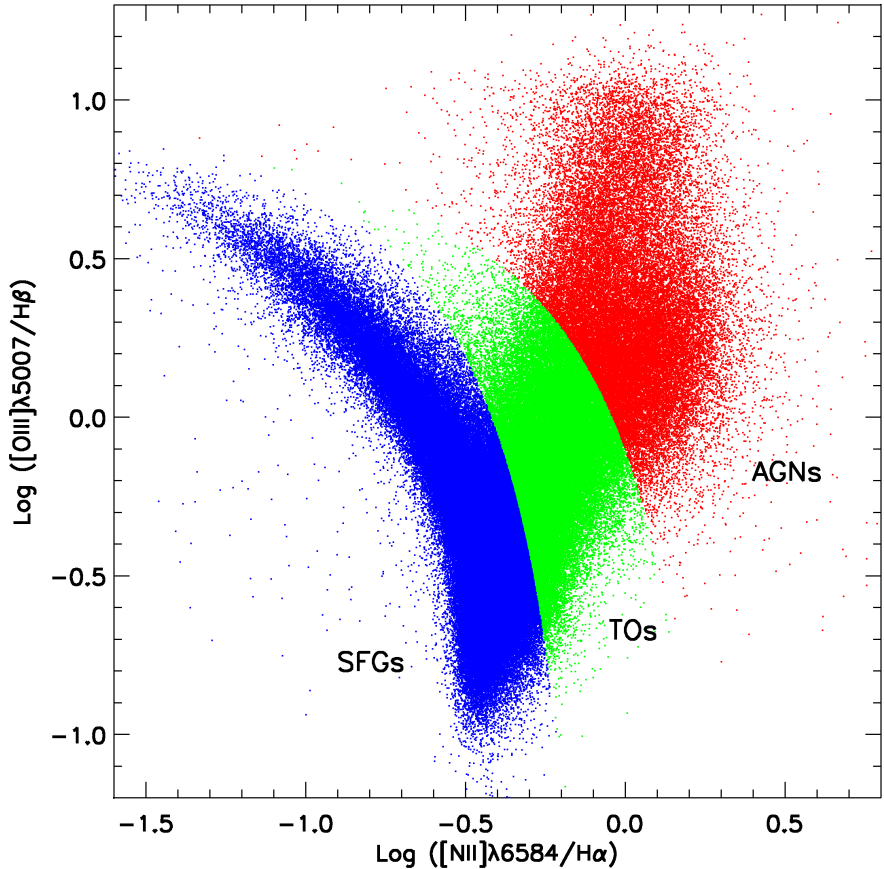}
  \caption{BPT-VO diagram for the NELGs. The criterion that separates SFGs from TOs is that of \citet{Kauffmann03} and the criterion that separates AGNs from TOs is that of \citet{Kewley01}. The SFGs are shown in blue, the TOs in green and the AGNs in red.}
  \label{fig:01}
\end{figure}

As we have mentioned in the previous section, our method allows us to obtain a very precise classification by activity types. In Figure~\ref{fig:01} we present our primary classification based on the most useful of all of the diagnostic diagrams \citep[the BPT-VO diagram;][]{Baldwin81,Veilleux87}. The separation criteria applied to distinguish between AGNs, TOs and SFGs are those proposed by \citet{Kewley01} and \citet{Kauffmann03}. Results for our classification are shown in Table~\ref{tab:01}. There we can see that the SFG type is the dominant activity in NELGs, with 53.5\% of the sample. The remaining 46.5\% are separated into 28.8\% TOs and 17.7\% AGNs. As we mentioned earlier, these results are consistent with the view that AGNs are rare at low redshift, and frequently accompanied by star formation activity in the circumnuclear regions, which greatly complicates their identification and/or study.

\begin{table}[!t]\centering
  \setlength{\tabnotewidth}{\columnwidth}
  \tablecols{4}
  % Stretch the space between table columns 
  \setlength{\tabcolsep}{2.0\tabcolsep}
  \caption{Number of galaxies in samples} \label{tab:01}
 \begin{tabular}{lccc}
    \toprule
                & SFGs         &  TOs         & AGNs         \\
    \midrule
Total           &  122751      & {\bf 66234}  & 40633       \\
SFG1s           &  {\bf 67513} &    ...       &  ...        \\
SFG2s           &  {\bf 55238} &    ...       &   ...       \\
{[OI]} Sy2s     &   ...        &    ...       & 12264       \\
{[OI]} LINERs   &   ...        &    ...       & 14102       \\
{[SII]} Sy2s    &   ...        &    ...       & 12911       \\
{[SII]} LINERs  &   ...        &    ...       & 19893       \\
Sy2s            &   ...        &    ...       & {\bf 16970} \\
LINERs          &   ...        &    ...       & {\bf 23663} \\
\bottomrule
  \end{tabular}\\
\label{tab:01}
\end{table}

It is important to emphasize that despite having determined the line ratios with great accuracy, the BPT-VO diagram still shows a continuous distribution. Therefore, the separation criteria applied in this diagram look somewhat arbitrary, allowing numerous crossed activity identifications or ambiguous classifications. This is another important difficulty in studying narrow-line AGNs, which could have serious consequences, in particular when the samples studied are small. 

\begin{figure}[!h]
  \includegraphics[width=\columnwidth]{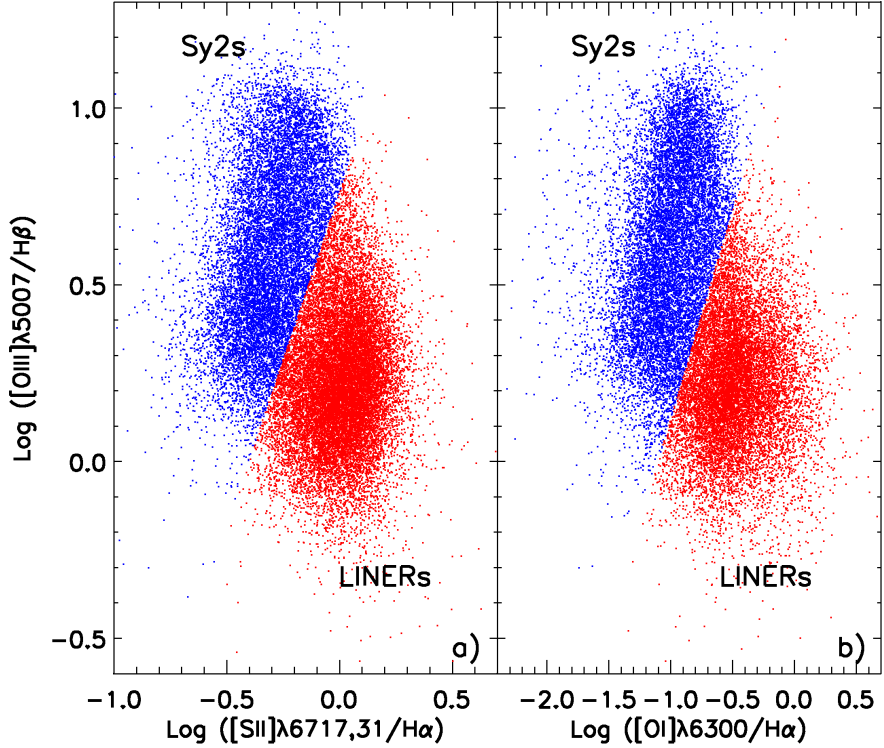}
  \caption{Diagnostic diagrams distinguishing between Sy2s and LINERs \citep{Kewley06}: a) [OIII]/H$\beta$ versus [SII]/H$\alpha$; b) [OIII]/H$\beta$ versus [OI]/H$\alpha$.}
  \label{fig:02}
\end{figure}

The difficulty becomes more obvious when one tries to differentiate between LINERs and Sy2s. No natural distinction appears in the BPT-VO diagram between these two activity types. In their study, \citet{Kewley06} suggested that LINERs and Sy2s may be distinguished based on differences in the [OI]$\lambda 6300$ and [SII]$\lambda\lambda6717, 6731$ emission-line intensities. This is shown in Figure~\ref{fig:02} where we reproduced  the analysis of these authors, applying their proposed criteria on our sample of AGNs to separate Sy2s from LINERs. The results are reported in Table~\ref{tab:01}. Note that only 65\% of the AGNs in our sample can be classified using [OI]$\lambda 6300$, which is a consequence of the intrinsically low intensity of this line in these galaxies (affecting LINERs more than Sy2s). On the other hand, even when using a lower-ionization line like [SII] only 81\% of the AGNs can be classified this way. Moreover, some galaxies change activity type passing from one diagram to another.

\begin{figure}[!h]
  \includegraphics[width=\columnwidth]{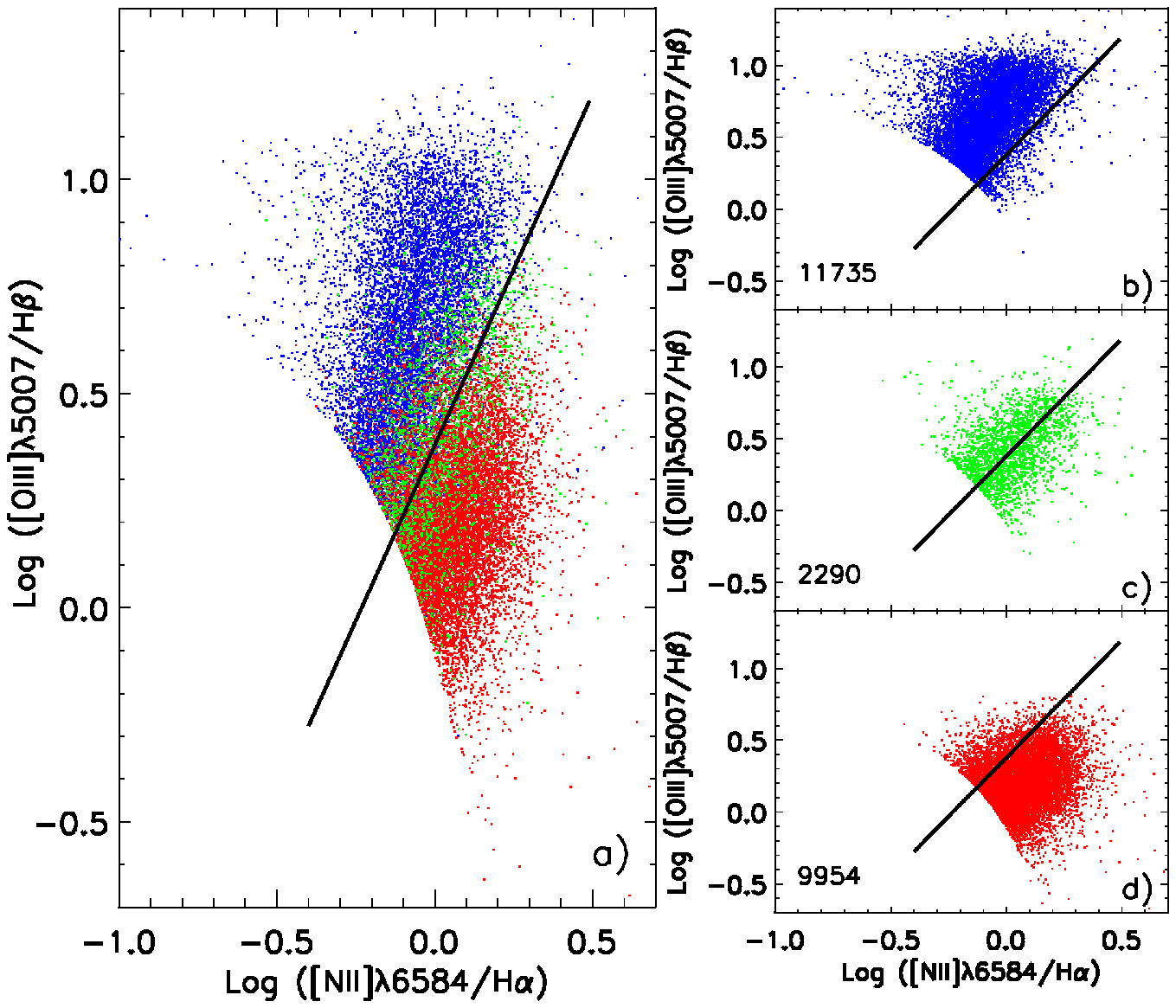}
  \caption{BPT-VO diagram for the NELGs classified as Sy2s and LINERs in Figure~\ref{fig:02}; in b) we show all the galaxies classified as Sy2s using two diagnostic diagrams; in d) we show those classified as LINERs; and in c) we show those that change type according to the diagnostic diagram. The continuous line separates this last sample in two equal-sized subsamples. This dividing line is reported in b) and d) and on the whole sample in a). }
  \label{fig:03}
\end{figure}

To be able to classify all the AGNs in our sample, we used the results of the analysis in Figure~\ref{fig:02} to establish a dividing line between the LINERs and Sy2s in the BPT-VO diagram. This is shown in Figure~\ref{fig:03}. In the right panel in~b) we have located in the BPT-VO the galaxies classified as Sy2s in Figure~\ref{fig:02}. In~d) we have located those classified as LINERs, and in~c) we have traced the position of the galaxies that change activity type depending on the diagnostic diagram used. In~a) we added the three subsamples together, which represent 53\% of the whole AGN sample. Based on this result, it is clear that we cannot eliminate completely the arbitrariness in the separation between the LINERs and Sy2s in any diagnostic diagram. As a compromise, we chose a dividing line that splits the sample of ambiguous types in~c) in two groups having the same number of galaxies and used this separation as our principal classification criterion in the BPT-VO diagram to distinguish between Sy2s and LINERs.  The separation corresponds to the following linear relation in Log:

\begin{equation}
{\rm Log} \left(\frac{[OIII]}{H\beta}\right) = 1.64 \times {\rm Log} \left(\frac{[NII]}{H\alpha}\right) + 0.38
\end{equation}

Note that our chosen criterion slightly favors Sy2s over LINERs (that is, in Figure~\ref{fig:03}, a slightly higher number of LINERs in d) than Sy2s in b) change activity classification). After applying this criterion to all the AGNs in our sample we count 58\% LINERs and 42\% Sy2s (see Table~\ref{tab:01}). 

The presence of star formation in narrow-line AGNs is now well recognized. However, this was not always the case in the past. At the time when the diagnostic diagrams were developed, the main assumption was that the right part of the continuous $\nu$-shaped distribution traced by the line ratios in the BPT-VO diagram was occupied by pure AGNs. With the introduction of the TO type the preoccupation now is slightly different. The current question is whether there could be AGNs among the SFGs. For instance, in Figure~\ref{fig:01} some SFGs show unusually high emission-line ratios of [NII]/H$\alpha$ as compared to what is observed in HII regions \citep[see][]{Coziol96}. This can be explained either by assuming an excess of excitation due to an AGN \citep[e.g.,][]{stasinska06}, or by assuming an excess of nitrogen abundance, related with the formation of massive bulges \citep{Coz99}. In \citet{Torres-Papaqui12} this problem was explored in great detail and it was concluded that, in general (except for a few ambiguous cases, as expected based on the ambiguity of selection criteria due to the continuous distribution of the NELGs in the BPT-VO diagram), there is no evidence of AGNs in the SFGs, in good agreement with the criterion proposed by \citet{Kauffmann03} to separate TOs from SFGs. Following \citet{Torres-Papaqui12}, we have separated our SFG sample in two. The SFG1 (55.2\% of the SFGs), are consistent with the standard view of late-type spirals having experienced constant star formation over their whole history. The SFG2s with an excess of nitrogen abundance form 44.8\% of the SFGs. These are massive early-type spiral galaxies (that is, spirals with massive bulges) that have experienced (or some are still experiencing) intense bursts of star formation during their formation.    

\begin{figure}[!h]
  \includegraphics[width=\columnwidth]{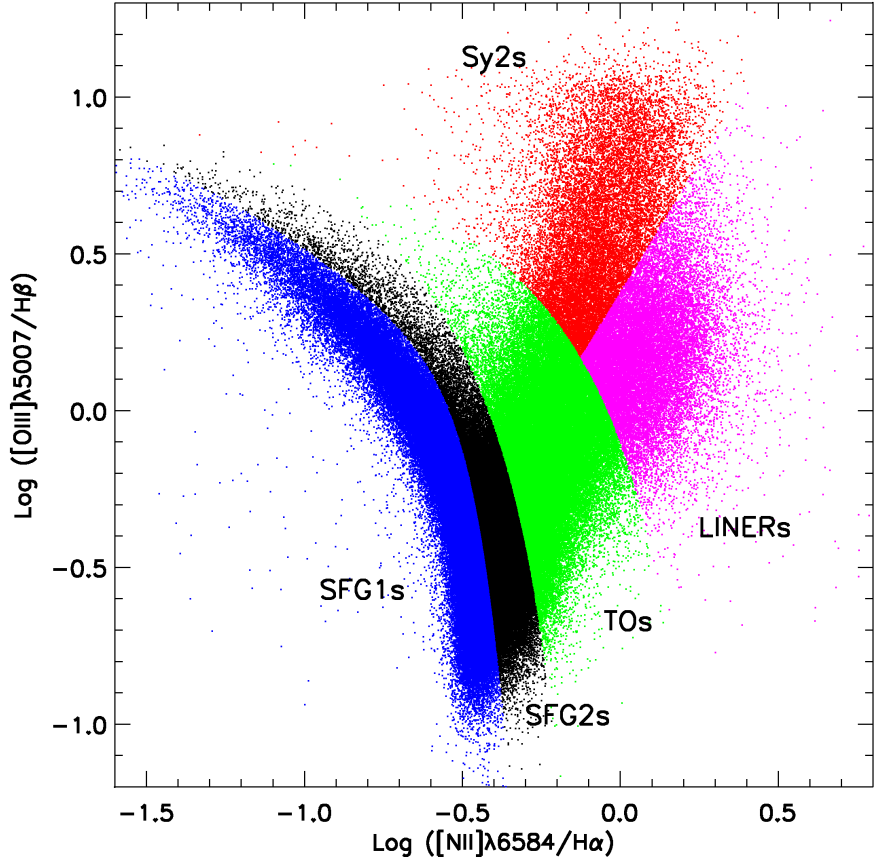}
  \caption{BPT-VO diagram defining the five different activity types used in our study: SFG~1 (blue), SFG~2 (black), TO (green), Sy2 (red) and LINER (magenta).}
  \label{fig:04}
\end{figure}

In Figure~\ref{fig:04} we show the final separations in activity type in the BPT-VO diagram, according to which all the galaxies received only one classification. The number of galaxies in each sample is identified in bold in Table~\ref{tab:01}. There are almost as many TOs as there are SFG1s. The number of SFG2s is also remarkably high, which indicates that powerful starburst galaxies are not rare in the nearby universe. The ``pure'' AGNs on the other hand are much less common. Among these, the LINERs are dominating over the Sy2s (this is despite the bias introduced by our classification criterion), which is consistent with previous surveys, showing that LINERs are very common in intermediate spiral galaxies in the field  \citep{Heckman80, Ho97}, and constitute the principal activity type of galaxies encountered in dense galactic environments \citep{Phillips86,Coziol98,Miller03,Martinez08,Martinez10}. Note that by extending the separation between LINERs and Sy2s to the TO region, more than 2/3 would classify as TO/LINER rather than TO/Sy2 galaxies. This result reflects the ambiguity found in the literature about the real nature of LINERs.   

\section{The Osterbrock test: first evidence for the presence of a SMBH in NELGs}
\label{sec:osterbrock}

\begin{figure}[!h]
  \includegraphics[width=\columnwidth]{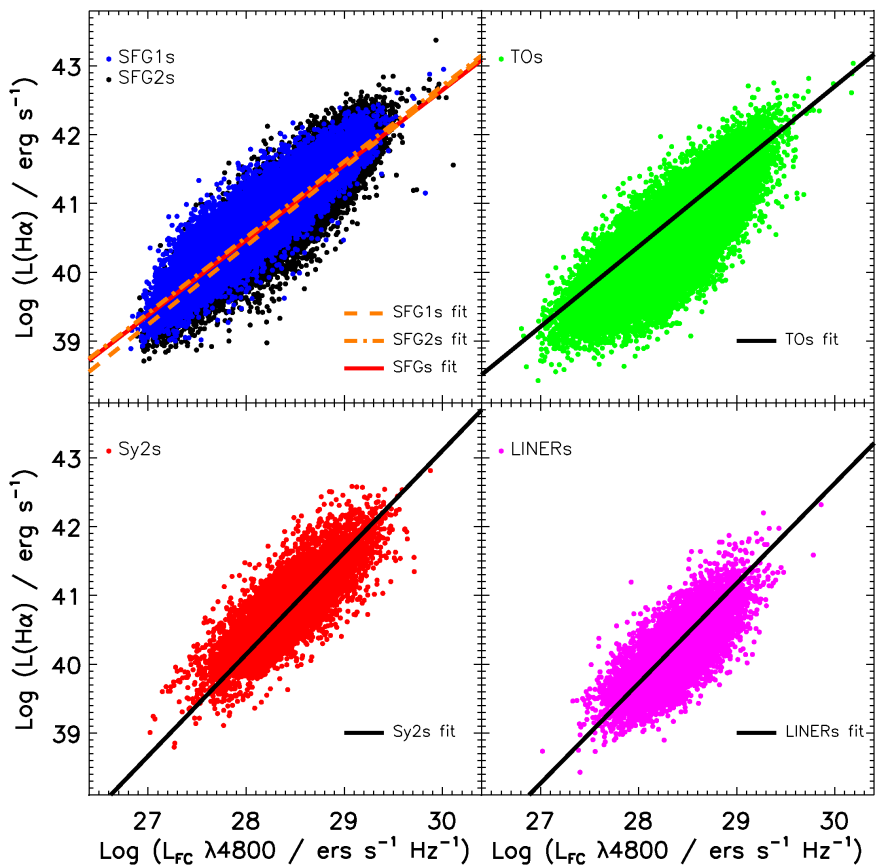}
  \caption{Results of the Osterbrock test for the NELGs separated by activity type: top-left, the SFG1s are plotted over the SFG2s (almost no difference in distribution); top-right, TOs; bottom-left, Sy2s and bottom-right, LINERs. The parameters of the linear fits are given in Table~\ref{tab:02}.}
  \label{fig:05}
\end{figure}

\begin{table}[!t]\centering
  \setlength{\tabnotewidth}{\columnwidth}
  \tablecols{5}
  % Stretch the space between table columns 
  \setlength{\tabcolsep}{2.0\tabcolsep}
  \caption{Parameters of linear fits and correlation coefficients} \label{tab:02}
 \begin{tabular}{lcccc}
    \toprule
    Sample & $\alpha$    &  $\beta$  & r$_s$ & r \\
    \midrule
    SFGs   & 1.09 $\pm$ 0.04 &  9.90 $\pm$ 0.35 &  0.91   & 0.91   \\
    SFG1s  & 1.10 $\pm$ 0.05 &  9.66 $\pm$ 0.34 &  0.88   & 0.89   \\
    SFG2s  & 1.14 $\pm$ 0.03 &  8.41 $\pm$ 0.34 &  0.93   & 0.93   \\
    TOs    & 1.16 $\pm$ 0.04 &  7.51 $\pm$ 0.37 &  0.77   & 0.79   \\
    Sy2s   & 1.48 $\pm$ 0.03 & -1.31 $\pm$ 0.03 &  0.81   & 0.82   \\
    LINERs & 1.45 $\pm$ 0.03 & -1.21 $\pm$ 0.03 &  0.75   & 0.75   \\
    \bottomrule
  \end{tabular}
\end{table}

The Osterbrock test is based on the principle that the source of ionization of an AGN is the continuum produced by the accretion of matter on the SMBH at its center. By comparing the luminosity of the line-emission with the luminosity of the continuum in the blue part of the spectrum, the AGNs should therefore follow a characteristic power law that can be easily distinguished from any relation produced by other ionizing sources, in particular O and B stars like in the SFGs (and possibly TOs). For our test we compare the luminosity in H$\alpha$ with the luminosity of the continuum at 4800\AA. The results are shown in Figure~\ref{fig:05} for the NELGs in our sample separated by activity type. In Table~\ref{tab:02} we give the parameters of the linear fits and the Spearman rank (r$_s$) and Pearson (r) coefficients of correlations. All the NELGs show strong linear correlations in Log of the kind:

\begin{equation}
{\rm Log} \left(\frac{{\rm L(H}\alpha)}{{\rm erg\, s^{-1}}}\right) = \alpha * {\rm Log} \left(\frac{\rm L_{FC}(4800\textrm{\AA})}{{\rm erg\, s^{-1}\, Hz^{-1}}}\right) + \beta
\end{equation}

Examining the parameters of the linear fits in Table~\ref{tab:02}, we see that the SFGs and TOs have similar $\alpha$ and $\beta$, suggesting their dominant source of ionization is similar. The LINERs and Sy2s also have similar $\alpha$ and $\beta$, which are significantly different from those determined for  the SFGs.  The trend is for the slope to be steeper in AGNs than in SFGs. That is, for the same luminosity in the emission line, the luminosity of the continuum is higher in an AGN than in a SFG. This is the main characteristic of AGNs. Therefore, we conclude that the ionization source in all the LINERs in our sample is the same as in the Sy2s. Consequently, both types are genuine AGNs \citep{Kewley06}. 

\begin{figure}[!h]
  \includegraphics[width=\columnwidth]{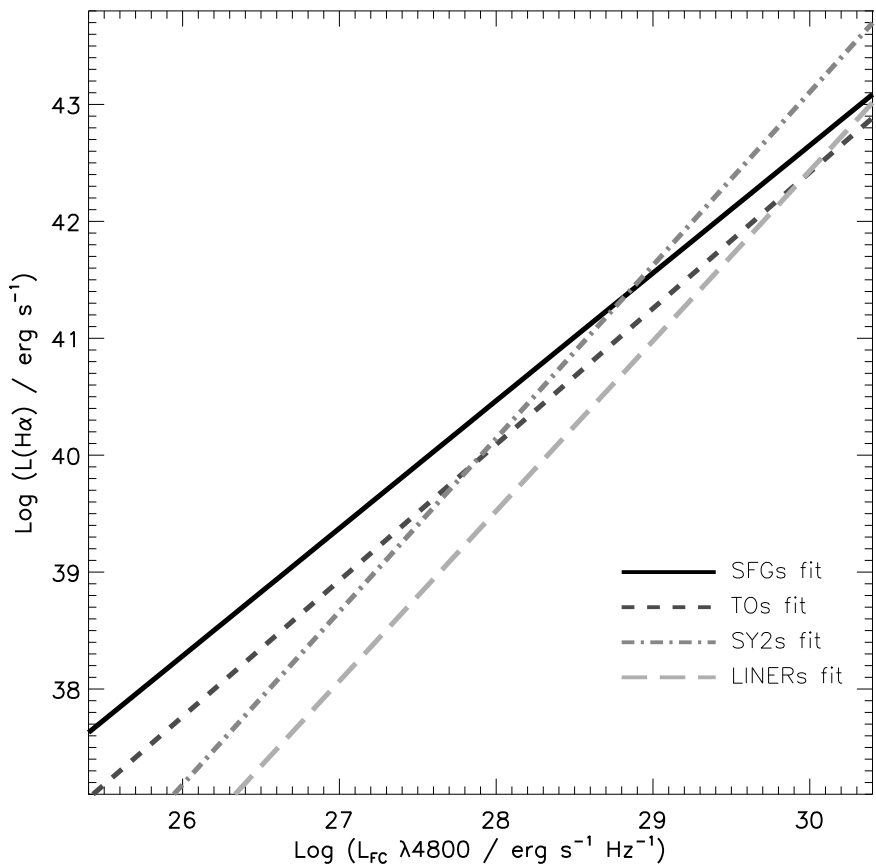}
  \caption{Comparison of the linear fits for the luminosity power law in NELGs having different activity types.}
  \label{fig:06}
\end{figure}

\begin{figure}[!h]
  \includegraphics[width=\columnwidth]{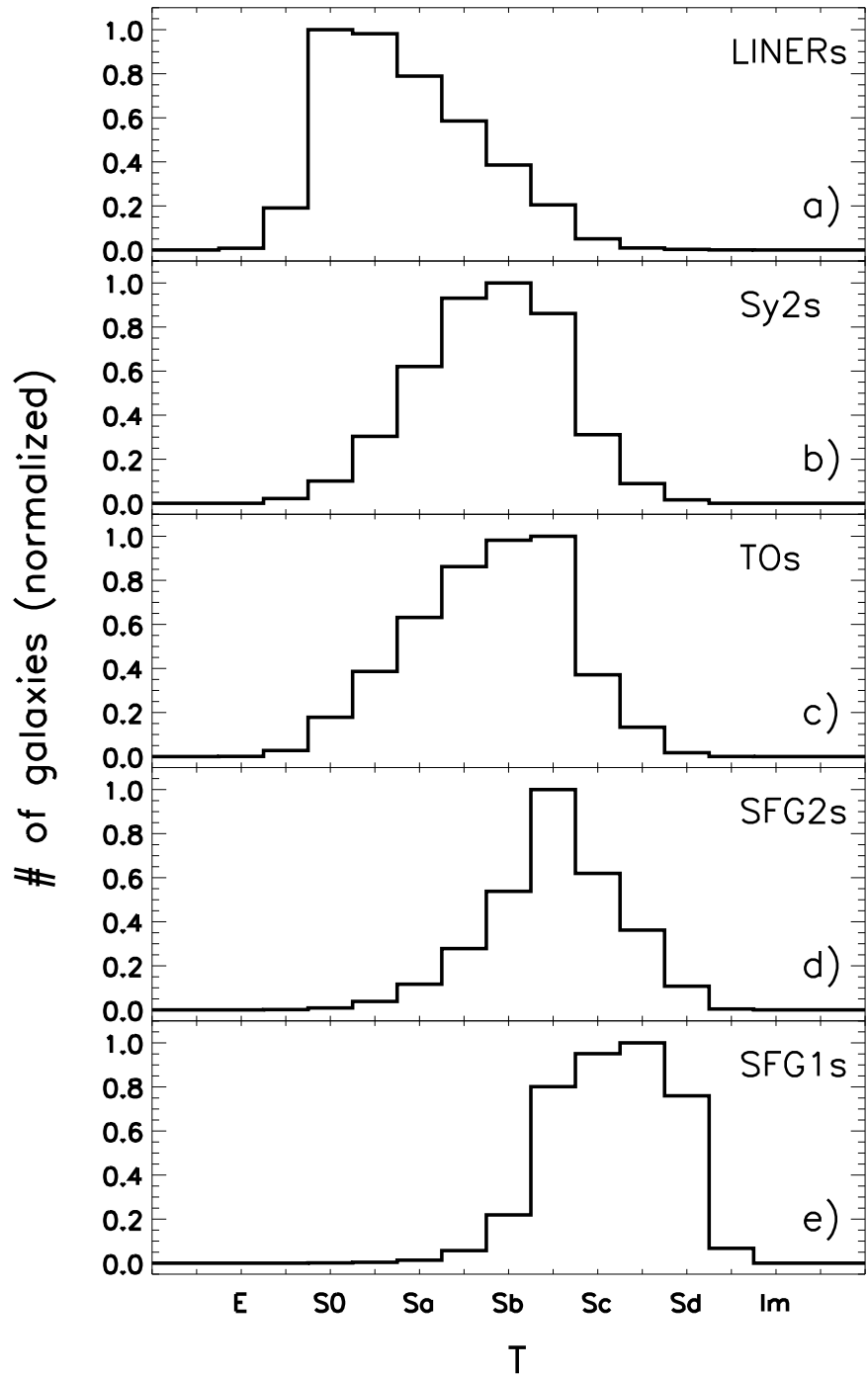}
  \caption{Morphology distributions of the NELGs with different activity types.}
  \label{fig:07}
\end{figure}

\begin{table}[!t]\centering
  \setlength{\tabnotewidth}{\columnwidth}
  \tablecols{6}
  % Stretch the space between table columns 
  \setlength{\tabcolsep}{1.2\tabcolsep}
  \caption{Parameters of linear fits and correlation coefficients according to activity type and morphology} \label{tab:03}
 \begin{tabular}{lccccc}
    \toprule
    Sample & T & $\alpha$    &  b  & r$_s$ & r \\
    \midrule
    SFG1s  & Sa $<$ T1 $\le$ Sb & 1.09 $\pm$ 0.06 &  9.70 $\pm$ 0.38 & 0.90 & 0.90 \\
    SFG1s  & Sb $<$ T2 $\le$ Sc & 1.15 $\pm$ 0.06 &  8.04 $\pm$ 0.34 & 0.93 & 0.93 \\
    SFG1s  & Sc $<$ T3 $\le$ Sd & 1.16 $\pm$ 0.06 &  7.90 $\pm$ 0.37 & 0.93 & 0.94 \\
    SFG2s  & Sa $<$ T1 $\le$ Sb & 1.14 $\pm$ 0.03 &  8.44 $\pm$ 0.34 & 0.83 & 0.84 \\
    SFG2s  & Sb $<$ T2 $\le$ Sc & 1.10 $\pm$ 0.03 &  9.55 $\pm$ 0.36 & 0.87 & 0.88 \\
    SFG2s  & Sc $<$ T3 $\le$ Sd & 1.14 $\pm$ 0.03 &  8.16 $\pm$ 0.38 & 0.91 & 0.92 \\
    TOs    & S0 $<$ T1 $\le$ Sa & 1.21 $\pm$ 0.05 &  6.24 $\pm$ 0.40 & 0.68 & 0.70 \\
    TOs    & Sa $<$ T2 $\le$ Sb & 1.11 $\pm$ 0.05 &  9.01 $\pm$ 0.37 & 0.77 & 0.79 \\
    TOs    & Sb $<$ T3 $\le$ Sc & 1.17 $\pm$ 0.05 &  7.27 $\pm$ 0.33 & 0.84 & 0.86 \\
    Sy2s   & S0 $<$ T1 $\le$ Sa & 1.37 $\pm$ 0.07 &  1.58 $\pm$ 0.07 & 0.74 & 0.75 \\
    Sy2s   & Sa $<$ T2 $\le$ Sb & 1.41 $\pm$ 0.07 &  0.47 $\pm$ 0.06 & 0.80 & 0.81 \\
    Sy2s   & Sb $<$ T3 $\le$ Sc & 1.39 $\pm$ 0.07 &  1.20 $\pm$ 0.07 & 0.86 & 0.86 \\
    LINERs & E  $<$ T1 $\le$ S0 & 1.43 $\pm$ 0.05 & -0.68 $\pm$ 0.09 & 0.72 & 0.73 \\
    LINERs & S0 $<$ T2 $\le$ Sa & 1.48 $\pm$ 0.06 & -2.07 $\pm$ 0.08 & 0.74 & 0.74 \\
    LINERs & Sa $<$ T3 $\le$ Sb & 1.49 $\pm$ 0.07 & -2.12 $\pm$ 0.08 & 0.80 & 0.80 \\
    \bottomrule
  \end{tabular}
\end{table}

Comparing the results of the linear fits in Figure~\ref{fig:06} a more complete and consistent picture can be drawn.  The LINERs are not only similar to the Sy2s by their ionization source, but also have significantly lower emission-line luminosity. In other words, the LINERs are the low-luminosity equivalent of the Sy2s, or better, since the LINERs are much more common than the Sy2s, the Sy2s are the high-luminosity equivalent of the LINERs. At the same time, we note that the TOs, although similar to the SFGs, present the same tendency of a steepening of the slope, which seems consistent with the standard interpretation of a TO as a galaxy having a double nature, that is, the TO type is a mixture of SFG and AGN. 

The Osterbrock test is unambiguously suggesting that the ionization source in the narrow-line AGNs is the continuum produced by the accretion of matter onto a SMBH at their center. Therefore, the power law should be the same, independent of the morphology. In Figure~\ref{fig:07} we show the results of our morphological classification for all the galaxies in our sample separated by activity types. There it can be seen that the morphologies change significantly from one activity type to the other. The SFG2s, for example, reside in spiral galaxies with slightly earlier type (Sbc) than the SFG1 (Sc/Scd). The TOs (Sab/Sb/Sbc) have even earlier type than the SFGs. They are, on the other hand, almost indistinguishable from the Sy2s. The LINERs have the earliest types of all (S0/S0a). 

To check if the linear relations varied with the morphology of the galaxies we have separated the morphology distributions in three different groups. The results of the linear fits in these groups are presented in Table~\ref{tab:03}. As expected, we observe no significant variation of the slope with the  morphology.  This suggests that the variations in stellar populations, being significantly older in the early-type galaxies, do not influence the linear relation between the ionized flux and continuum. This shows that the slope of the linear relation depends only on the activity type. In the SFGs and TOs young massive stars, which are common in all star-forming galaxies, are dominating the continuum in the blue part of the spectrum, explaining the similar slopes. Similarly in the Sy2s and LINERs, the same characteristic slope suggests a common phenomenon different from star formation, which is the accretion of matter onto a SMBH at their center. The LINERs and Sy2 are similar types of AGNs, differing only by their intrinsic line-emission luminosity. 

\section{Concentration of mass in the center of galaxies and FWHM: other evidence in favor of a SMBH at the center of NELGs}
\label{sec:VMA}

\begin{table}[!t]\centering
  \setlength{\tabnotewidth}{\columnwidth}
  \tablecols{5}
  % Stretch the space between table columns
  \setlength{\tabcolsep}{0.8\tabcolsep}
  \caption{VMA and FWHM in NELGs separated according to activity type} \label{tab:04}
 \begin{tabular}{lccccccccc}
    \toprule
              &  \multicolumn{4}{c}{Log VMA} & &\multicolumn{4}{c}{Log 
FWHM(H$\alpha$)} \\
              &  1Q & Median & Mean & 3Q & &1Q & Median & Mean & 3Q \\
    \midrule
    SFG1s     &  9.37  &  9.56 &  9.59 &  9.78 & & 2.40 &  2.41 &  2.43 & 2.45 \\
    SFG2s     &  9.62  &  9.81 &  9.84 & 10.06 & & 2.43 &  2.47 &  2.48 & 2.52 \\
    TOs       &  9.80  & 10.04 & 10.04 & 10.27 & & 2.47 &  2.53 &  2.54 & 2.59 \\
    Sy2s      &  9.85  & 10.11 & 10.09 & 10.34 &  &2.51 &  2.58 &  2.59 & 2.66 \\
    LINERs    & 10.00  & 10.23 & 10.22 & 10.45 & & 2.53 &  2.60 &  2.61 & 2.68 \\
    \bottomrule
  \end{tabular}
\end{table}

\begin{figure}[!h]
  \includegraphics[width=\columnwidth]{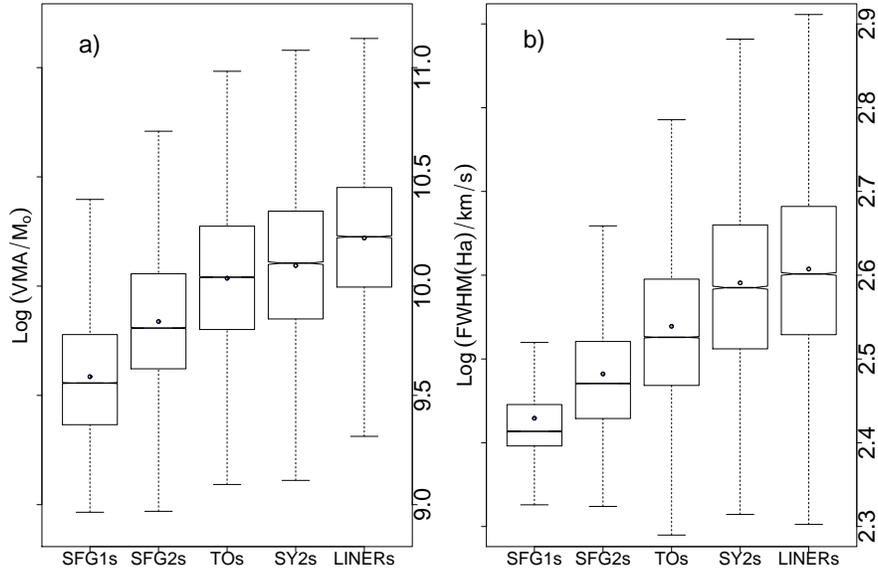}
  \caption{Box-whisker plots for VMA and FWHM. The extremely small notches (barely visible) are due to the large sizes of the samples, and reflect the significant differences in medians of these samples. The dots correspond to the means.}
  \label{fig:08}
\end{figure}

The Osterbrock test for the NELGs is consistent with the presence of an accreting SMBH at the center of  the Sy2s and LINERs. We have already shown that the LINERs reside in galaxies having the earliest morphological types of all the NELGs in our sample, which suggests they have generally massive bulges. However, the TOs show no important morphological differences from the Sy2s, which suggests that the presence of a massive bulge is not sufficient to explain why a SMBH becomes active in a galaxy. A more restrictive constraint is the gravitational binding energy \citep[or gravitational potential energy; see][]{Hopkins07b,Aller07,Coziol11}. Assuming the formation of a SMBH follows the formation process of galaxies by gravitational collapse, then galaxies developing such objects at their nucleus should have naturally higher binding energy than those without it, and this would be irrespective of the morphology of the galaxy. By definition, a higher gravitational binding energy implies that more mass is being concentrated at the center of these galaxies. Having at hand the velocity dispersion in the aperture of the spectra, $\sigma_{ap}$, we can easily estimate the virial mass within the aperture (VMA), and use this as a proxy for the binding energy: the higher the mass within the aperture the higher the binding energy. 

We define the VMA as: 
 \begin{equation}
{\rm VMA} = \frac{R \sigma_{ap}^2}{G}
\end{equation}
where $R$ is the physical radius projected by the aperture. The box-whisker plots in Figure~\ref{fig:08}a reveal significant differences of the VMA as a function of the activity type.  The values describing the statistical distributions are reported in Table~\ref{tab:04}. The median and mean for the LINERs and Sy2s are significantly higher than for the SFGs and TOs. An ANOVA statistical test confirms the significance of these differences at a 95\% confidence level (see appendix A). From our analysis we find that the VMA gradually increases from the SFG1s and SFG2s, to the TOs and Sy2s, culminating in the LINERs.   

\begin{figure}[!h]
  \includegraphics[width=\columnwidth]{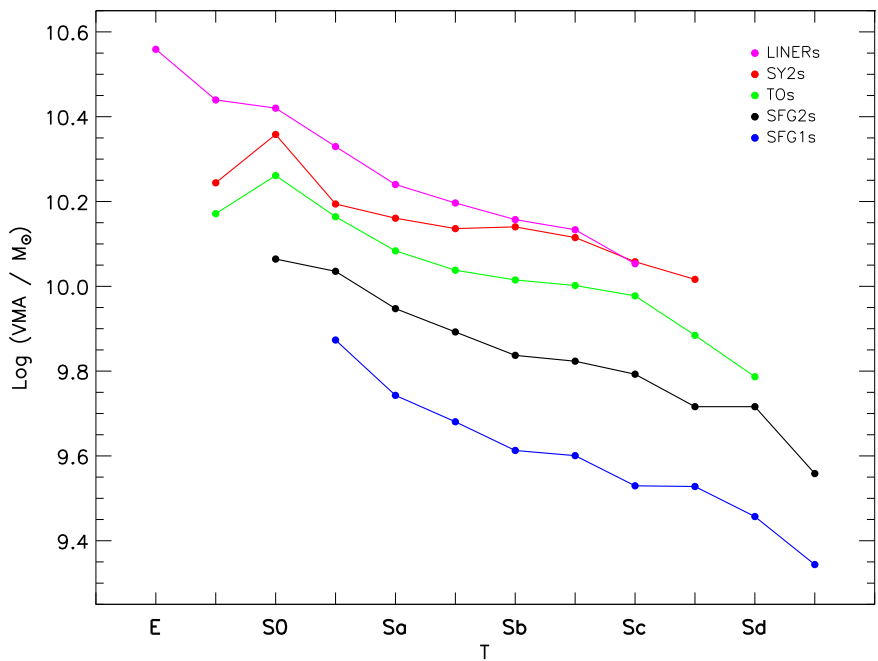}
  \caption{Median VMA as a function of morphology, for galaxies with different activity types.}
  \label{fig:09}
\end{figure}

In Figure~\ref{fig:09} we show the median VMA for the NELGs with different activity types and different morphologies. The VMA is always higher in the AGNs. The fact that this characteristic is independent of the morphology of the galaxies is important. It clearly shows that the VMA does not depend on the mass of the bulge, as we suggested. The AGNs have higher VMA because they have higher gravitational binding energies, which is due to the formation of a SMBH at their centers. 

Note also the different trends in Figure~\ref{fig:09}: the LINERs have significantly higher VMA than the Sy2s in early-type galaxies, but the difference between the two decreases in the later types, while the Sy2s differ more significantly from the TOs in the late types than in the early types. In general, the TOs have a VMA $ \ge 10^{10}$ M$_{\odot}$, significantly higher than the SFG2s. Since they have physical characteristics comparable to those of the Sy2s, differing possibly only in their higher star-formation rates, it seems plausible to assume that a SMBH is also present at the center of these galaxies. 

\begin{figure}[!h]
  \includegraphics[width=\columnwidth]{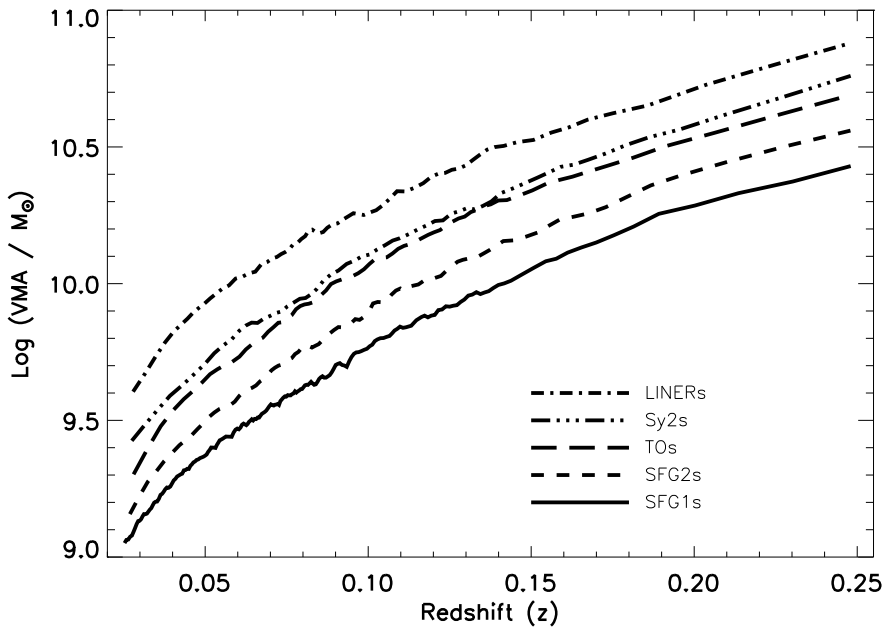}
  \caption{Median VMA as a function of redshift for galaxies with different activity types.}
  \label{fig:10}
\end{figure}

According to its definition (cf. Eq.~3), the VMA varies linearly with the projected aperture size, and consequently it varies with the redshift. We show in Figure~\ref{fig:10} the median VMA as a function of redshift for galaxies having different activity types. It can be seen that at any redshift the VMA is always higher in the AGNs. Note that we do not observe much difference between the TOs and Sy2s in this diagram, which reinforces the idea that TOs also have a SMBH at their center. From this comparison it is clear that AGNs have a characteristically high VMA, independent of the the projected aperture size. That is, going from high to low redshift the projected aperture size decreases, but the VMA of an AGN keeps being higher than in other activity types. At the limit where the projected aperture size become unresolved, this behavior is consistent with the presence of an unresolved massive object at the center of all the AGNs. 

\subsection{Relation of VMA with the FWHM}

\begin{figure}[!h]
  \includegraphics[width=\columnwidth]{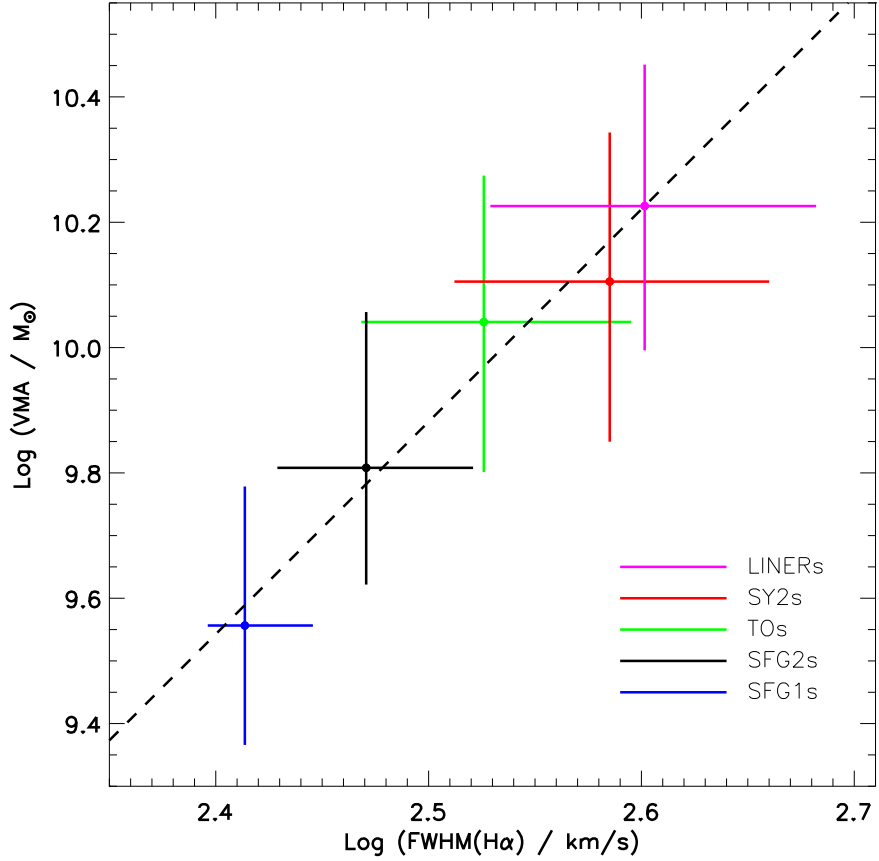}
  \caption{Relation of FWHM with VMA. The points correspond to the medians for different activity types and the bars cover the first and third quartiles.}
  \label{fig:11}
\end{figure}

In the past \citep[e.g.,][]{Osterbrock89}, it was fairly well accepted that one defining characteristic of AGNs is that their emission-lines have large Full Width at Half Maximum (FWHM). Narrow-line AGNs are no exception. For instance, Sy2s have permitted and forbidden lines of similar widths in the range 500 km s$^{-1}$, which is significantly broader than in a typical SFG where the FWHM is of the order of  200 km s$^{-1}$. Here we show that this characteristic is directly related with the VMA, and a consequence for AGNs of having an unresolved SMBH at their center. In Figure~\ref{fig:08}b, we show the box-whisker plots for the FWHM of the H$\alpha$ line in galaxies having different activity types. We observe that the FWHM increases in the same way as the VMA. This can also be observed in Table~\ref{tab:04}. As for the VMA, an ANOVA statistical test confirms that the differences are statistically significant (see appendix A). In Figure~\ref{fig:11} we report the VMA and FWHM medians, first and third quartiles for the NELGs with different activity types. The dashed line is a linear relation that quantifies the gradual increase of  VMA with the FWHM:
\begin{equation}
{\rm Log} \left(\frac{\rm VMA}{\rm M_{\odot}}\right) = (3.39\pm 0.03) \times {\rm Log}\left(\frac{{\rm FWHM(H}\alpha)}{{\rm km\ s}^{-1}}\right) + (1.4\pm 0.4)
\end{equation}
Based on this result, we see that the TOs with FWHM $ \ge 10^{2.5}$\ km s$^{-1}$ are more similar to Sy2s than to SFGs. 

The reason why AGNs have characteristically large FWHM is not well understood. In terms of dynamical Doppler effects, 
broad-line profiles translate into high-speed gas motions, either rotational and/or turbulent. What is clear, on the other hand, is that the high level of energy necessary to produce these movements cannot come from thermal sources, namely massive starburst winds or SN explosions \citep{BNW90}, which anyway seems difficult to assume in LINERs where the star formation activity may be at its lowest level. Our result suggests that this energy is gravitational, related with the high binding energy produced by the formation of a SMBH at the center of the galaxies. 

\section{Discussion: masses of the SMBHs and accretion rates in narrow-line AGNs}
\label{sec:discussion}

\begin{figure}[!h]
  \includegraphics[width=\columnwidth]{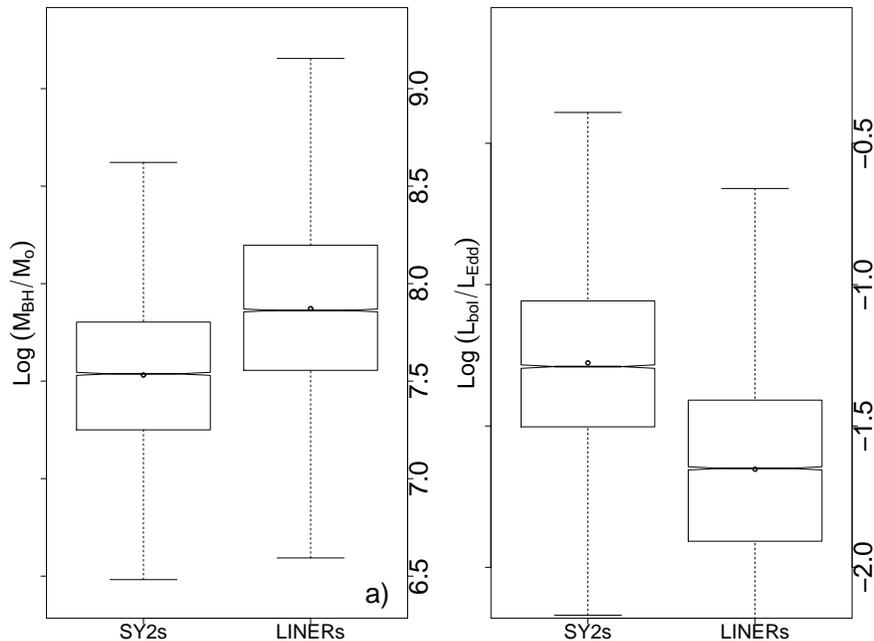}
  \caption{Box-whisker plots for the masses of the SMBHs and accretion rates in the narrow-line AGNs. The notches, which are barely visible due to the large sizes of the two samples, reflect the significant differences in medians between the samples. The dots represent the means.}
  \label{fig:12}
\end{figure}

In the two previous sections we presented new evidence for the presence of SMBHs at the center of NELGs. Added to those already recognized in the literature (in particular the BPT-VO diagnostic diagram) this evidence strongly supports the interpretation that narrow-line AGNs, like the Sy2s and LINERs, are genuine. Within this interpretation, we now discuss how the standard AGN model, the accretion of matter onto a SMBH, may explain the differences between the different types. For this discussion, the masses of the SMBHs were estimated following \citet{Gultekin09} and the accretion rates correspond to L$_{Bol}$/L$_{Edd}$, where the bolometric luminosity is obtained using the approximate relation L$_{Bol} = 9\lambda$L$_{\lambda}$(5100\AA) \citep{Kaspi00}.  The continuum is evaluated at a different wavelength than before in order to be able to compare our results with those of Broad-Line AGNs \citep[BLAGNs; in particular][]{Peterson04,Peterson05}. 

\begin{table}[!t]\centering
  \setlength{\tabnotewidth}{\columnwidth}
  \tablecols{5}
  % Stretch the space between table columns
  \setlength{\tabcolsep}{0.8\tabcolsep}
  \caption{SMBH masses and accretion rates in narrow-line AGNs} \label{tab:05}
 \begin{tabular}{lccccccccc}
    \toprule
              &  \multicolumn{4}{c}{Log(M$_{BH}$ / M$_{\odot}$)} & &
\multicolumn{4}{c}{Log(L$_{bol}$/L$_{Edd}$)} \\
              &  1Q & Median & Mean & 3Q & &1Q & Median & Mean & 3Q \\
    \midrule
    Sy2s       & 7.25 & 7.54 & 7.53 & 7.80 & &-1.50 & -1.29 & -1.28 & -1.06 \\
    LINERs     & 7.55 & 7.86 & 7.87 & 8.20 & &-1.92 & -1.65 & -1.65 & -1.41 \\
    \bottomrule
  \end{tabular}
\end{table}

\begin{figure}[h]
  \includegraphics[width=\columnwidth]{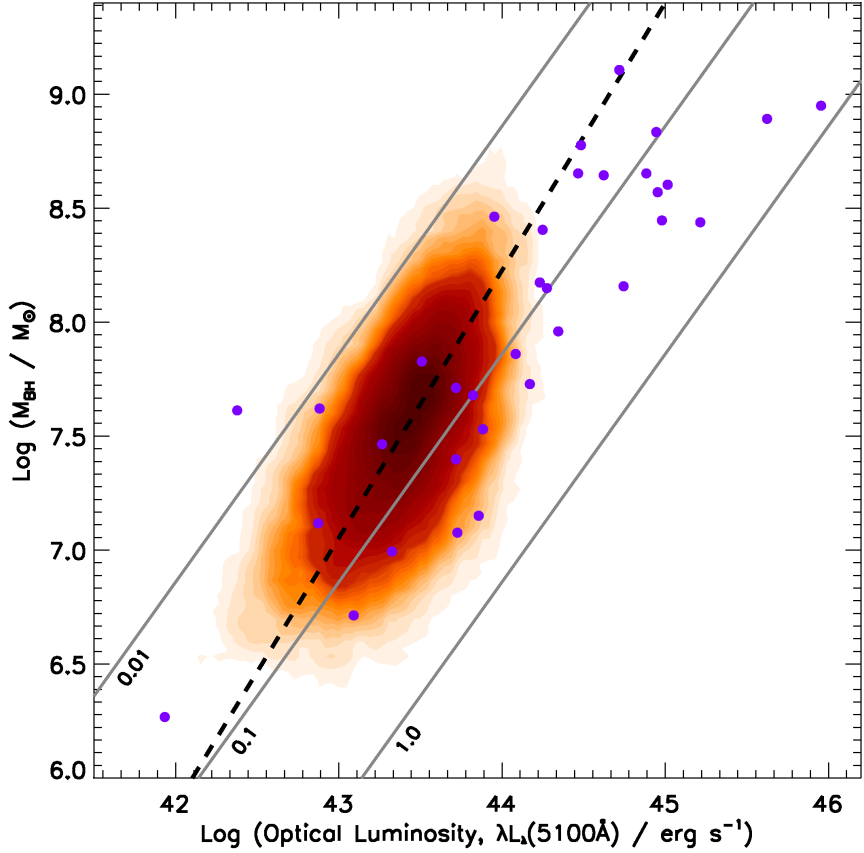}
  \caption{Mass of SMBH vs. accretion luminosity for the Sy2s. We show the data as shaded density contours; the darker the region the more numerous the data points. The individual data points correspond to the sample of BLAGNs as previously studied by \citet{Peterson05}. The continuous lines are different accretion rates, L$_{Bol}$/L$_{Edd}$. The dashed line is a linear fit to our data.}
  \label{fig:13}
\end{figure}

In Figure~\ref{fig:12} we draw the box-whisker plots for the masses of the SMBHs  and their accretion rates. The notches once again are barely visible due to the large sizes of our samples, also emphasizing the statistically significant difference of the medians. The values describing the distributions are reported in Table~\ref{tab:05}. The statistical significance for the differences of the means was established using ANOVA tests (see appendix A). Our results are comparable to those obtained before by \citet{Kewley06}, who used different methods to obtain the masses of the SMBHs and accretion rates. Our main conclusion is consequently the same as these authors: the LINERs have more massive SMBHs that accrete at lower rates than the Sy2.

In Figure~\ref{fig:13} we compare the masses of the SMBHs and the accretion luminosity as we measured in the Sy2s with those estimated in BLAGNs observed in reverberation by \citet{Peterson04} and studied in \citet{Peterson05}. In general the Sy2s have lower-mass SMBHs than the BLAGNs. In Figure~\ref{fig:13} the Sy2s seem also to be accreting at lower rates than the BLAGNs, i.e., below 0.1. Note that for comparable masses of SMBH the accretion rates we have determined are in good agreement with those estimated by \citet{Peterson04}. Our much larger sample however allows us to observe a new trend for the more massive SMBHs in the Sy2s to have lower accretion rates. If we trace a linear correlation over the data the slope for the Sy2s is above 1, 1.17 $\pm$ 0.04 to be exact, with correlation coefficients of 0.6 (both Spearman rank and Pearson, with probabilities of obtaining the same values by chance practically zero). This correlation suggests that as the SMBH increases in mass, its luminosity, or accretion rate, decreases.  

\begin{figure}[!h]
  \includegraphics[width=\columnwidth]{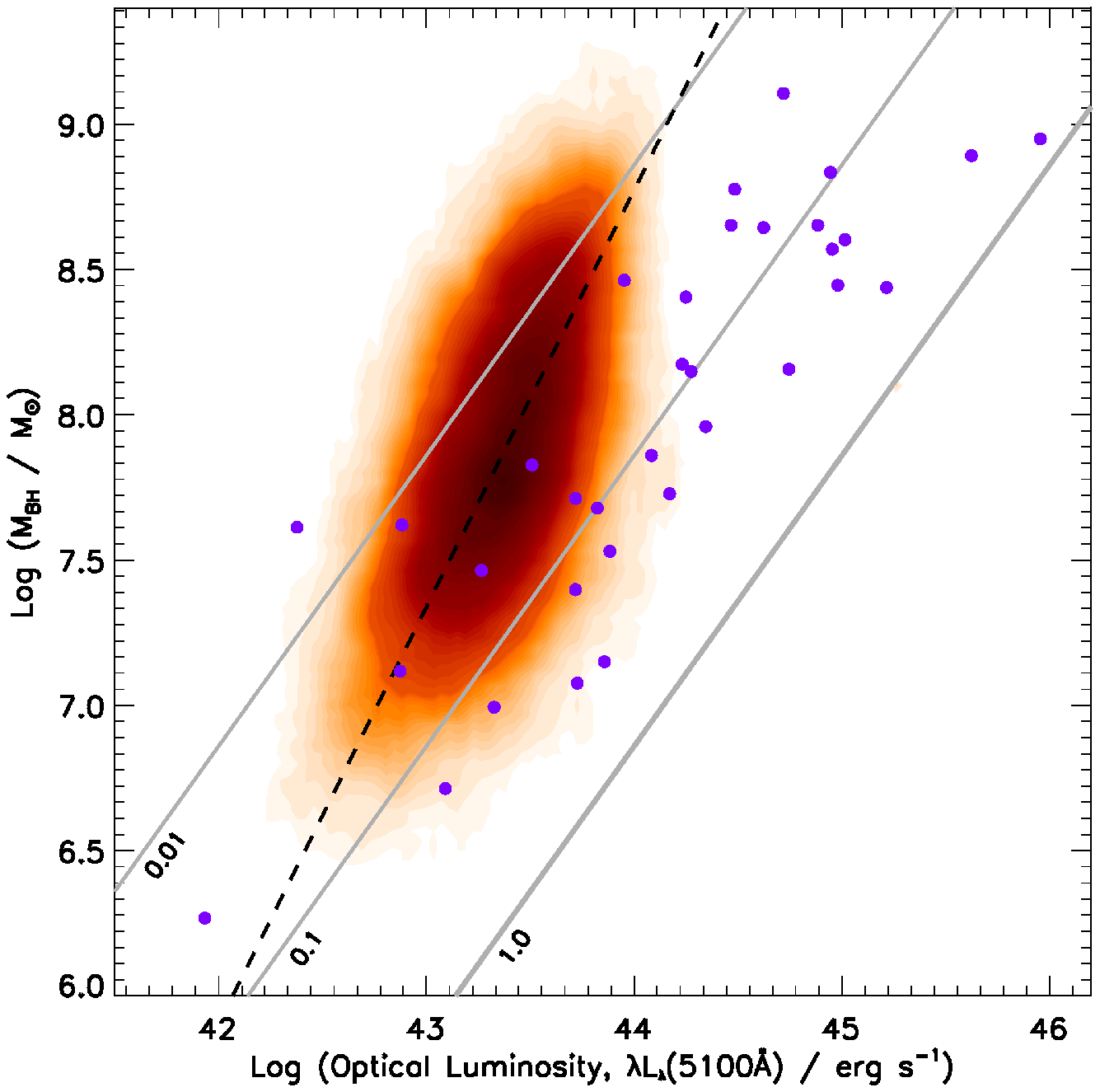}
  \caption{Same as Figure~\ref{fig:13} for the LINERs.}
  \label{fig:14}
\end{figure}

In Figure~\ref{fig:14} we compare the LINERs with the BLAGNs. Here the differences are much more important. In the LINERs some of the SMBHs reach almost $10^9$\ M$_\odot$, comparable to the most luminous BLAGNs. However, they accrete at much lower rates, around 0.01. Like for the Sy2, but with a steeper slope of 1.43 $\pm$ 0.04, a linear correlation (also with correlation coefficient 0.6) suggests that as the SMBH increases in mass the accretion rate decreases. 

\begin{figure}[!h]
  \includegraphics[width=\columnwidth]{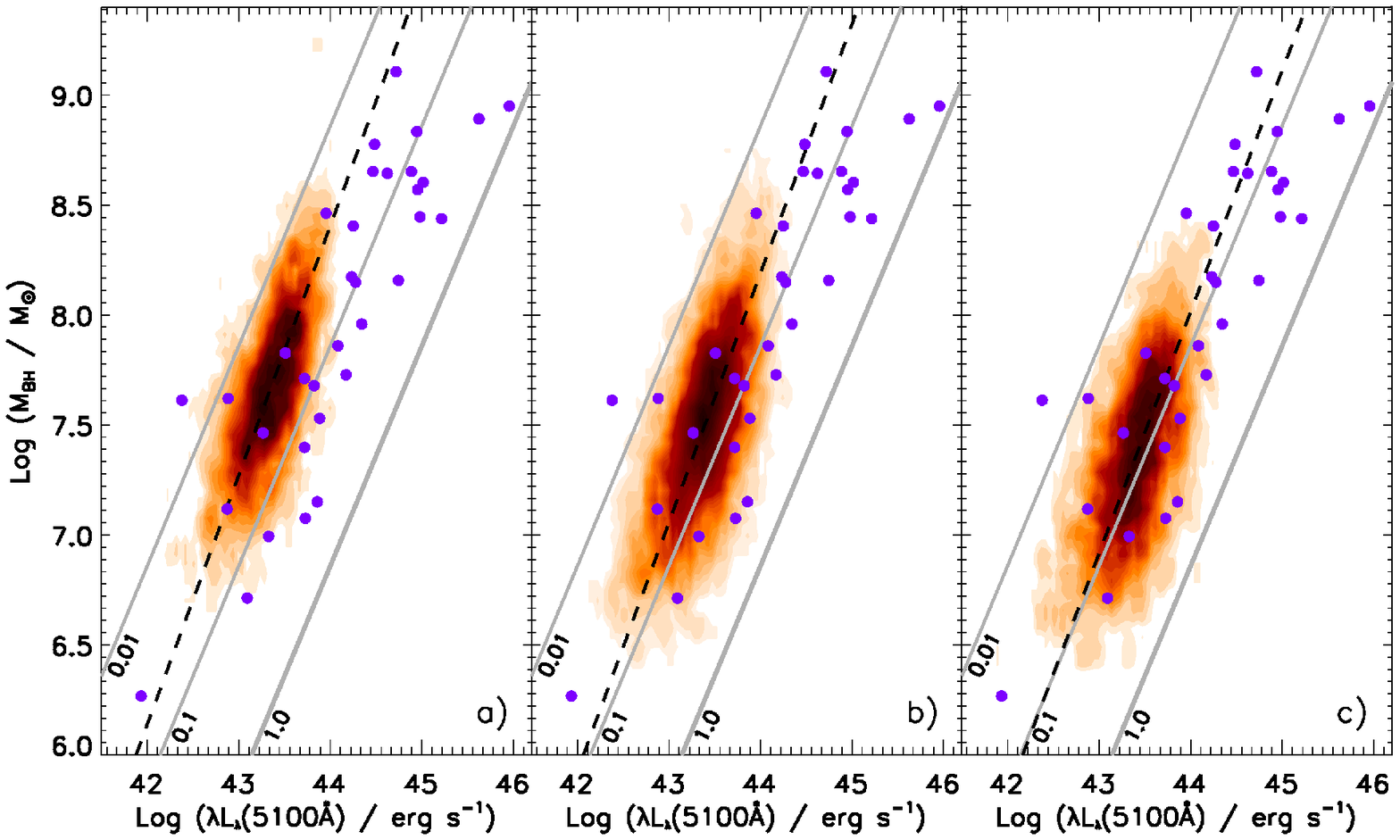}
  \caption{Same as Figure~\ref{fig:13} separating the sample of Sy2s in three different morphology groups: a) early-type T1; b) intermediate-type  T2; c) late-type T3. The definition of the groups can be found in Table~\ref{tab:03}.}
  \label{fig:15}
\end{figure}

\begin{figure}[!h]
  \includegraphics[width=\columnwidth]{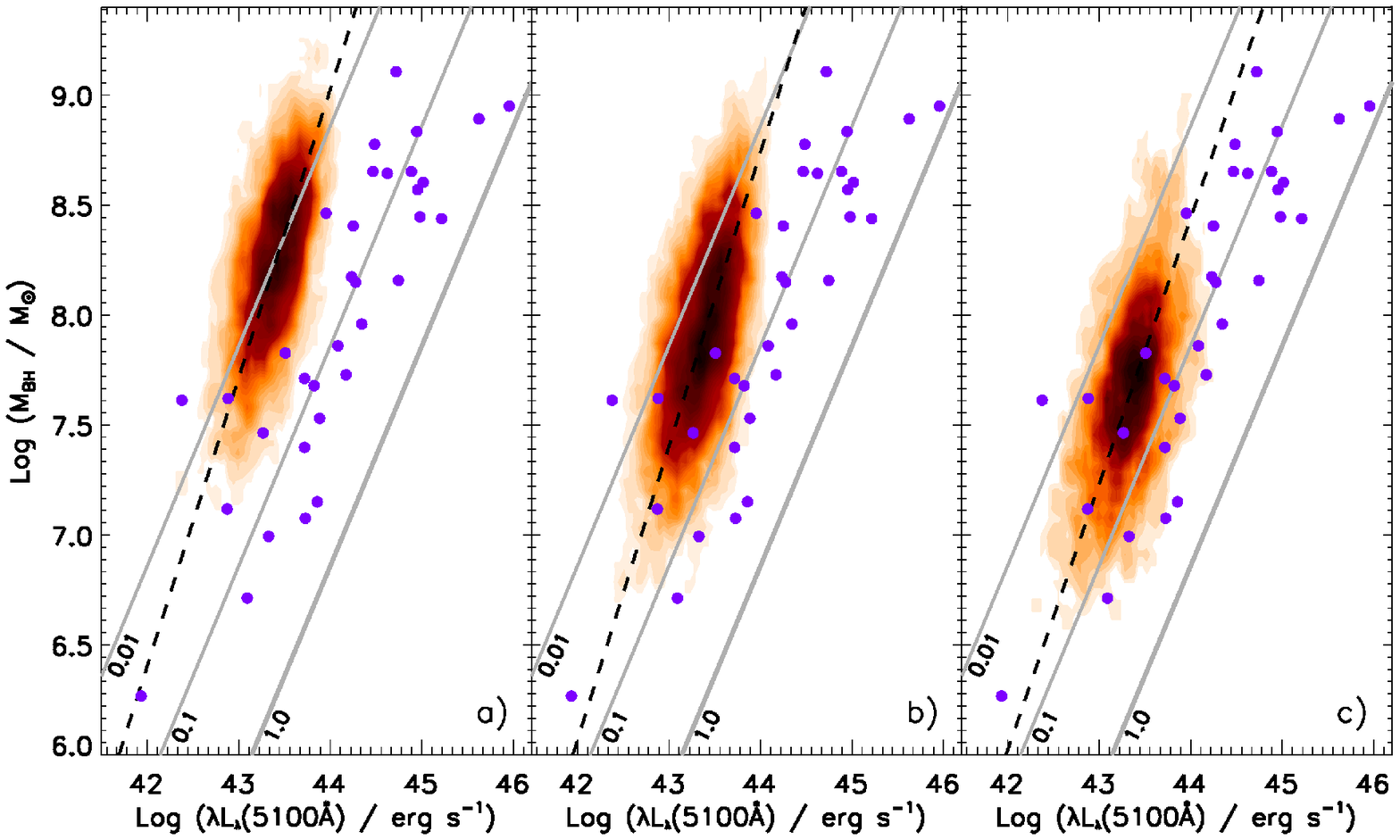}
  \caption{Same as Figure~\ref{fig:15} for the LINERs.}
  \label{fig:16}
\end{figure}

In parallel with the variation of accretion rate with the SMBH mass, we also observe a variation in morphology. In Figure~\ref{fig:15} for the Sy2s we show the SMBH mass as a function of accretion luminosity in three different morphological groups: in a) the early-type group T1,  in b) the intermediate-type group T2, and in c) the late-type group T3. The definition of these morphological groups are given in Table~\ref{tab:03}. We see systematic changes, the SMBH masses increasing from the late types to the early types, and the accretion rates decreasing in the same sense. Moreover, the slope of the linear correlation increases from 1.09 $\pm$ 0.04 in the group T3 to 1.13 $\pm$ 0.03 in the T2 and T1 groups. The same trend is observed for the LINERs separated in morphological groups in Figure~\ref{fig:16}. There is a clear augmentation in the mass of the SMBH passing from the late types to the early types, concurrent with a significant decrease in the accretion rates. The slope of the linear correlation also steepens passing from 1.21 $\pm$ 0.03 in the T3 to 1.34 $\pm$ 0.05 in the T2 and 1.31 $\pm$ 0.04 in the T1. 

In Table~\ref{tab:06} we give the characterictic statistics for the masses of the SMBH as they varied with the morphoplogcal types and in  Table~\ref{tab:07} we give the characterictic statistics for the luminosities. For the Sy2s, on average a small dimming in luminosity by 18\% is observed going from the T3 to the T1, but not in the LINERs where the luminosity increases by 11\%. The differences in luminosity between the LINERs and Sy2s are also marginal: the T1 LINERs are 15\% more luminous than the T1 Sy2s, while the T3 are 13\% dimmer. However, at the same time we see a constant increase in the SMBH mass from the T3 to the T1 in both the Sy2s (the T1 are 53\% more massive than the T3) and the LINERs (the T1 are 75\% more massive than the T3), and we see an increase by 70-85\% comparing the LINERs with the Sy2s by morphologies. Therefore, for almost the same bolometric luminosity the mass of the SMBH increases significantly with the morphological type, which translates into a decrease in accretion rate (or accretion efficiency) with the mass of the SMBH. 

 \begin{table}[!t]\centering
  \setlength{\tabnotewidth}{\columnwidth}
  \tablecols{5}
  % Stretch the space between table columns
  \setlength{\tabcolsep}{0.8\tabcolsep}
  \caption{Masses of SMBHs in narrow-line AGNs having different morphological types}
\label{tab:06}
 \begin{tabular}{lccccccccc}
    \toprule
 Act.  &  Morpho. & \multicolumn{4}{c}{Log (M$_{BH}$ / M$_{\odot}$)} \\
       &          & 1Q & Median & Mean  & 3Q \\
    \midrule
       & T1   &  7.44 &  7.67   &  7.68 &  7.92 \\
SY2s   & T2   &  7.26 &  7.52   &  7.52 &  7.78 \\
       & T3   &  7.11 &  7.38   &  7.38 &  7.65 \\
    \midrule
       & T1   &  8.04 &  8.27   &  8.26 &  8.50 \\
LINERs & T2   &  7.66 &  7.92   &  7.93 &  8.20 \\
       & T3   &  7.44 &  7.68   &  7.68 &  7.90 \\
    \bottomrule
  \end{tabular}
\end{table}

\begin{table}[!t]\centering
  \setlength{\tabnotewidth}{\columnwidth}
  \tablecols{5}
  % Stretch the space between table columns
  \setlength{\tabcolsep}{0.8\tabcolsep}
  \caption{Luminosity of narrow-line AGNs having different morphological types}
\label{tab:07}
 \begin{tabular}{lccccccccc}
    \toprule
 Act.  &  Morpho. & \multicolumn{4}{c}{Log
($\lambda$L$_{\lambda}$(5100\AA) / erg s$^{-1}$)} \\
    \midrule
       &          & 1Q & Median & Mean  & 3Q  \\
    \midrule
       & T1   & 43.16 &  43.39  & 43.36 & 43.58 \\
Sy2s   & T2   & 43.20 &  43.43  & 43.41 & 43.63 \\
       & T3   & 43.21 &  43.45  & 43.43 & 43.66 \\
    \midrule
       & T1   & 43.25 &  43.43  & 43.42 & 43.60 \\
LINERs & T2   & 43.20 &  43.41  & 43.39 & 43.60 \\
       & T3   & 43.17 &  43.38  & 43.37 & 43.56 \\
    \bottomrule
  \end{tabular}
\end{table}

The above trends are consistent with some pattern in evolution \citep[e.g.][]{Kelly10}. As the bulge of the galaxies grows (due to star formation) the SMBH grows in mass through accretion. The bulges stop growing when no more stars form, which means the amount of available gas in these galaxies is already too low, and possibly also the amount of gas to be accreted by the SMBH, explaining the lower accretion luminosity. Note that for the Sy2s with accretion rates near 0.1 L$_{Edd}$ in the group T3, the SMBH may continue to grow significantly in mass, and if this accretion is accompanied by intense star formation, the bulge would also grow, eventually transforming the morphology of the galaxy to an earlier type. Consistent with the relation between the SMBH mass and the mass of the bulge, our observations suggest the Sy2s are still very much active in building both, their bulges through star formation concurrently with their SMBH throuh accretion.

Different from the Sy2s, the accretion rates in the LINERs are much lower, which suggests that the LINERs may already be at, or very near, the end of their formation process, which is also consistent with their earlier morphologies.  Therefore, the LINERs may have been more active in the past. Considering the similarity in binding energy and variation with morphology (Figure~\ref{fig:09}) this higher luminosity phase for the LINERs may have been similar to that in the Sy2s. Consequently, it may be that in each group of morphology the Sy2s will eventually evolve with time into something similar to a LINER.   

\section{Conclusions}

We have shown that the NELGs, identified as LINERs and Sy2s, trace similar power laws between the luminosity in emission and the luminosity in the continuum, the LINERs differing from the Sy2s only by their lower luminosity. This result is consistent with the standard model of AGN, which states that the principal source of ionizing photons in AGNs is the accretion of matter onto a SMBH at their center. Also consistent with the presence of a SMBH at the center of NELGs, we have shown that the VMA, that is, the mass concentrated at the center of the galaxies, is always higher in the AGNs than in galaxies showing a higher level of star formation, like the TOs and SFGs. The fact that this is independent of the morphology of the galaxies suggests this characteristic is not merely related to the formation of the bulge, but to a more fundamental aspect of the formation of the galaxies. This higher concentration of mass in the AGNs is related with a higher gravitational binding energy due to the formation of a compact massive object, a SMBH, at their center.  We have also found a strong correlation between the VMA and the FWHM, which suggests that the energy necessary to produce the large movement of gas in the AGNs is gravitational in nature, related with the binding energy of the SMBH at their center. 

We conclude that the standard AGN model depicts in an extremely consistent way the narrow-line AGN phenomenon observed at low redshift. According to this model SMBHs form at the center of most galaxies, growing, through active star formation, with the mass of their bulges \citep[see][and references therein]{Cattaneo01}. Therefore, one should not be surprised to find so ample evidence of AGNs with star formation. The TOs are possibly one good example. As the mass of the bulge and of the SMBH grow the galaxy transforms into an earlier morphological type. Eventually star formation fades away, leaving only a narrow line AGN (a LINER), where a SMBH still accretes matter but at a lower rate than before. According to our findings, it is very possible that all the narrow-line AGNs passed by a phase of higher activity in the past, which could suggest that most LINERs were Sy2s sometime in the past, and most probably, the Sy2s will eventually end their life as LINERs. 

The standard AGN model explains very well the $\nu$-shaped distribution traced by the NELGs in the BPT-VO diagram. This distribution reflects the  accumulation of mass at the center of galaxies through the formation of a SMBH and a massive bulge. On the left branch of the $\nu$-shaped distribution, we find galaxies with small binding energies. These galaxies form most of their stars in a disk and do not develop massive bulges.  The relatively small-mass SMBH that form in their nucleus do not accrete much matter either, which explains why we do not detect them. Our own galaxy is possibly of this kind. On the right branch of the $\nu$-shaped distribution, going from left to right, we find galaxies with increasing binding energies, and consequently more massive bulges and actively accreting SMBHs. The TOs, Sy2s and LINERs in this scenario may be related to different mass scales in the formation of a SMBH. Alternatively they may represent different evolutionary phases which would explain why the exact boundaries between the activity types are difficult to delineate. The LINERs in particular may represent the end products of this evolutionary process. 

Within the standard AGN model interpretation the Sy2s and LINERs seem to show clear evidence of evolution (within each type): as the SMBH grows in mass its accretion rate decreases.  This would suggest that the activity phase of the SMBH in narrow-line AGNs is of longer duration than usually assumed based on quasar models, varying over a period of time of the order of a Gyr or more \citep[e.g.,][]{Kewley06},  larger than the time necessary to form their bulges. In fact, in the narrow-line AGNs we may still see SMBHs growing in mass by accretion, in parallel with the host galaxies building their bulges through star formation (the Sy2s, and possibly also the TOs, are good examples). This longer activity timescale however may not apply to the more luminous AGNs like the quasars, suggesting possibly different accretion regimes \citep[e.g.,][]{SmallBlandford92,Hopkins06b}. In particular, we see no evidence of negative feedback from the activity of the SMBH, as suggested to regulate the activity of the quasars \citep[see][and reference therein]{Page12}. This model would not explain the variation of bulge characteristics observed in the narrow-line AGNs, passing from the TOs to the Sy2s and LINERs, which is more consistent with a positive feedback.

\section*{Acknowledgments}
T-P acknowledges PROMEP for support grant 103.5-10-4684 and DAIP-UGto (0065/11) and I. P.-F. acknowledges the postdoctoral fellowship grants 145727 and 170304 from CONACyT, M\'exico. The authors acknowledge an anonymous referee for important comments and suggestions that help us improve the quality of our article. The SDSS is managed by the Astrophysical Research Consortium (ARC) for the Participating Institutions. The Participating Institutions are: the American Museum of Natural History, Astrophysical Institute Potsdam, University of Basel, University of Cambridge (Cambridge University), Case Western Reserve University, the University of Chicago, the Fermi National Accelerator Laboratory (Fermilab), the Institute for Advanced Study, the Japan Participation Group, the Johns Hopkins University, the Joint Institute for Nuclear Astrophysics, the Kavli Institute for Particle Astrophysics and Cosmology, the Korean Scientist Group, the Los Alamos National Laboratory, the Max-Planck-Institute for Astronomy (MPIA), the Max-Planck-Institute for Astrophysics (MPA), the New Mexico State University, the Ohio State University, the University of Pittsburgh, University of Portsmouth, Princeton University, the United States Naval Observatory, and the University of Washington.

\appendix

\section{Results of statistical test}

The statistical framework of the tests we used is based on a new parametric ANOVA model introduced by \citet{Hothorn08} and developed for the R software\footnote{http://CRAN.R-project.org} by \citet{Herberich10}. The max-t test does the simultaneous pairwise comparisons of means under control of the family-wise error rate, which is the probability of falsely rejecting the initial hypothesis (i.e., finding a significant difference among the means of any two groups in the data set even though there is actually no difference present). The test takes into account possible heteroscedasticity and the unequal sizes of the groups.

We present the results of the tests in the form of simultaneous confidence intervals for all pairwise comparisons of group means. Confidence intervals including zero indicate no statistically significant differences. Confidence intervals near zero suggests some level of similarity, and the farther from zero they are the more significant the differences. Note the smallness of some of the confidence intervals. This is due to the very large number of data used in each bin, which makes the tests extremely significant.

\begin{figure}[!h]
  \includegraphics[width=\columnwidth]{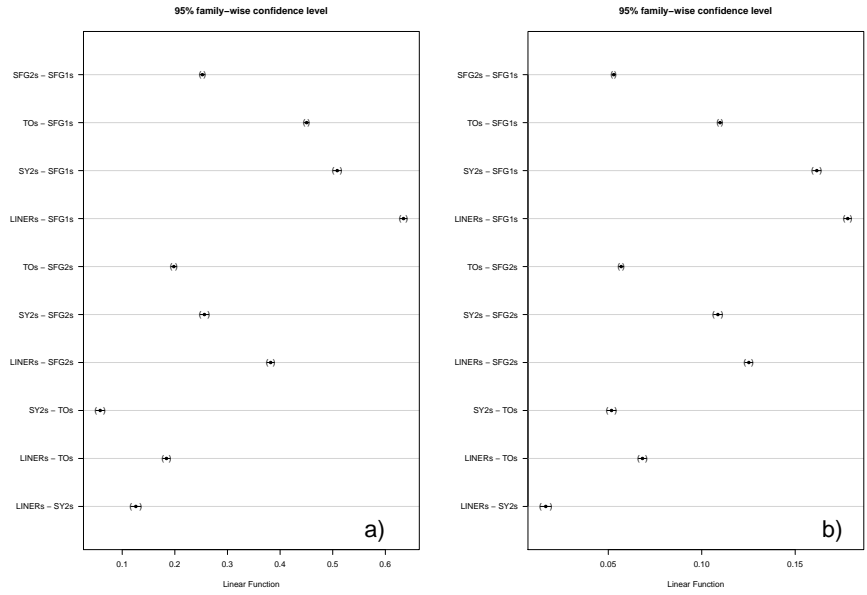}
  \caption{The confidence intervals associated with Figure~\ref{fig:08} in Section~\ref{sec:VMA}; a) VMA; b) FWHM of the H$\alpha$.}
  \label{fig:17}
\end{figure}

In Figure~\ref{fig:17} we show the confidence intervals associated with Figure~\ref{fig:08} (Section~\ref{sec:VMA}) where we compare the variations of physical parameters in NELGs with different activity types. In a) we see the VMA increasing in the sense, SFG1s $\Rightarrow$ SFG2s $\Rightarrow$ TOs $\Rightarrow$ Sy2s $\Rightarrow$ LINERs,  with a significant minor difference between the TOs and Sy2s; b) the FWHM increases in the same direction, with a minor difference between the Sy2s and LINERs. 

\begin{figure}[!h]
  \includegraphics[width=\columnwidth]{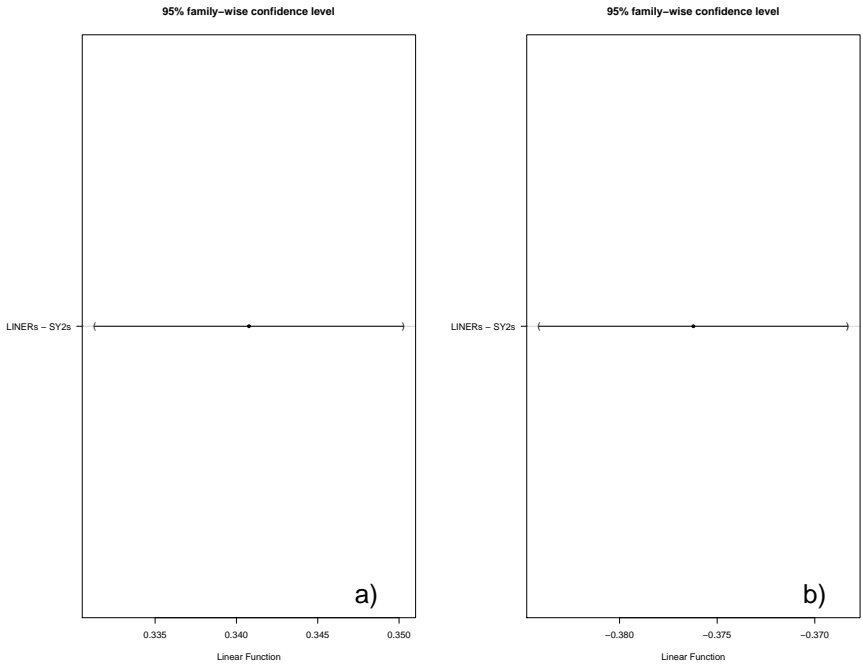}
  \caption{The confidence intervals associated with Figure~\ref{fig:12} in Section~\ref{sec:discussion}:  a) the masses of the SMBHs; b) the accretion rates.}
  \label{fig:18}
\end{figure}

In Figure~\ref{fig:18} we show the confidence intervals associated with Figure~\ref{fig:12} (Section~\ref{sec:discussion}), where we compare in a) the masses of the SMBHs  and in b) the accretion rates. We see that LINERs have significanlty more massive SMBHs that accrete at lower rates than the Sy2.

\end{document}